\documentclass[12pt]{iopart}

\usepackage{iopams,graphicx} 
\begin{document}

\title[Transition from Baryon- to Meson-Dominated Freeze Out]
{Transition from Baryon- to Meson-Dominated Freeze Out -- Early Decoupling around 30 $A$ GeV?}

\def\da{$^{a}$}
\def\uct{$^{b}$}
\def\wro{$^{c}$}

\author{H Oeschler\da, J Cleymans\uct, K Redlich\wro, S Wheaton\da$^,$\uct}

\address{
\da \mbox{Technische Universit\"at Darmstadt, D-64289 Darmstadt, Germany}\\
\uct \mbox{
University of Cape Town, Rondebosch 7701, South Africa}\\
\wro \mbox{Institute of Theoretical Physics, University of Wroc\l
aw, Pl-45204 Wroc\l aw, Poland}\\
}

\ead{h.oeschler@gsi.de}

\begin{abstract}
The recently discovered sharp peak in the excitation function of
the K$^+/\pi^+$ ratio around 30 $A$ GeV in relativistic heavy-ion
collisions is discussed in the framework of the Statistical Model.
In this model, the freeze-out of an ideal hadron gas changes from
a situation where baryons dominate to one with mainly mesons. This
transition occurs at a temperature $T$ = 140 MeV and baryon
chemical potential $\mu_B$ = 410 MeV corresponding to an energy of
$\sqrt{s_{\rm NN}}$ = 8.2 GeV. The calculated maximum in the
K$^+/\pi^+$ ratio is, however, much less pronounced than the one
observed by the NA49 Collaboration. The smooth increase of the
K$^-/\pi^-$ ratio with incident energy and the shape of the
excitation functions of the $\Lambda/\pi^+$, $\Xi^-/\pi^+$ and
$\Omega^-/\pi^+$ ratios all exhibiting maxima at different
incident energies, is consistent with the presently available
experimental data. The measured K$^+/\pi^+$ ratio exceeds the
calculated one just at the incident energy when the freeze-out
condition is changing.

We speculate that at this point freeze-out might occur in a
modified way. We discuss a scenario of an early freeze-out which
indeed increases K$^+/\pi^+$ ratio while most other particle
ratios remain essentially unchanged. Such an early freeze-out is
supported by results from HBT studies.
\end{abstract}

\maketitle

\section{Introduction}

The NA49 Collaboration  has recently  performed a series of
measurements of Pb-Pb collisions at 20, 30, 40, 80 and 158 $A$ GeV
beam energies \cite{Gazdzicki,NA49,Lambda-NA49}. Combining these
results with measurements at lower beam energies from the
AGS~\cite{pi-AGS,Ahle_1999,Ahle_2000,Lambda-AGS,Ahmad,Klay} they
reveal a sharp variation with beam energy in the
$\Lambda/\left<\pi\right>$, with $\left<\pi\right>\equiv
3/2(\pi^++\pi^-)$,
 and in the
K$^+/\pi^+$ ratios. Such a strong variation with energy does not
occur in pp collisions and therefore indicates a major difference
in heavy-ion interactions. A strong variation with energy of the
$\Lambda/\left<\pi\right>$ ratio has been predicted on the basis
of  arguments put forward in~\cite{Gorenstein}. It has also been
suggested recently in Ref.~\cite{Stock} that this might be a
signal of the critical point in the QCD phase diagram at high
baryon density.

In this paper we explore  another, less spectacular, possibility
for the sharp maximum. The arguments for this will be presented in
two steps. In step (i) the Statistical Model of a hadron gas is
studied. For this purpose, we investigate various quantities along
the freeze-out curve~\cite{1gev,JC_SQM2006} as a function of
$\sqrt{s_{\rm NN}}$. It is shown that the entropy of the hadronic
gas is dominated at the lower energies by baryons and at the
higher ones by mesons. The transition occurs just at the energy
where the sharp maximum has been observed. In step (ii) we
speculate whether at these incident energies the freeze-out might
happen earlier. The consequences are discussed and arguments based
on HBT results are presented to support this idea.

\section{Maximum relative strangeness content
 in heavy-ion collisions around 30 $A$ GeV}

The experimental data from heavy-ion collisions show that the
K$^+/\pi^+$ ratio rises from SIS up to AGS. It is larger for AGS
than at the highest CERN-SPS energies
\cite{NA49,1gev,Ahle,Dunlop,bearden} and decreases even further at
RHIC \cite{Star}. This behavior is of particular interest as it
could signal the appearance of new dynamics for strangeness
production in high-energy collisions. It was  even conjectured
that this property could indicate   an energy    threshold  for
quark-gluon-plasma formation in relativistic heavy-ion
collisions~\cite{Gazdzicki}.

In the following we analyze the energy dependence of strange to
non-strange particle  ratios in the framework of a hadronic
Statistical Model. In the whole  energy range, the hadronic yields
observed in heavy-ion collisions resemble a system in chemical
equilibrium characterized by two parameters, the temperature $T$
and the baryon chemical potential $\mu_B$. These two parameters
define the ``freeze-out curve'' which seems determined by the
condition of fixed average energy/particle $\simeq$ 1 GeV
\cite{1gev,JC_SQM2006}. As the beam energy increases in the energy
domain of SIS and AGS, $T$ rises and $\mu_B$ decreases slightly.
Above AGS energies, $T$ exhibits only a moderate change and
converges to its maximal value in the range of 160 to 180 MeV,
while $\mu_B$ is strongly decreasing.

Rather than studying the K$^+/\pi^+$ ratio, we use the ratios of
 strange to non-strange particle
multiplicities (Wroblewski factor)~\cite{wroblewski} defined as $
\lambda_s \equiv {2\bigl<s\bar{s}\bigr>\over \bigl<u\bar{u}\bigr>
+ \bigl<d\bar{d}\bigr>} $ where the  quantities in angular
brackets refer to the number of newly formed quark-antiquark
pairs, i.e.~it excludes all quarks that were present in the target
and the projectile nuclei.

Applying the Statistical Model to particle production in heavy-ion
collisions calls for the  use of the canonical ensemble to treat
the number of strange particles, particularly  for data in the
energy range from SIS up to AGS \cite{CLE99}. The calculations for
Au-Au and Pb-Pb collisions  are performed using a canonical
correlation volume~\cite{CLE99}. The quark content used in the
Wroblewski factor is determined at the moment of {\it {chemical
freeze-out}}, i.e.~from the hadrons and, especially, hadronic
resonances before they decay. This ratio is thus not an easily
measurable observable unless one can reconstruct all resonances
from the final-state particles.  The results are shown in Fig.~1
left as a function of $\sqrt{s_{\rm NN}}$.

\begin{figure}[h]
\begin{center}
\includegraphics*[width=7.5cm]{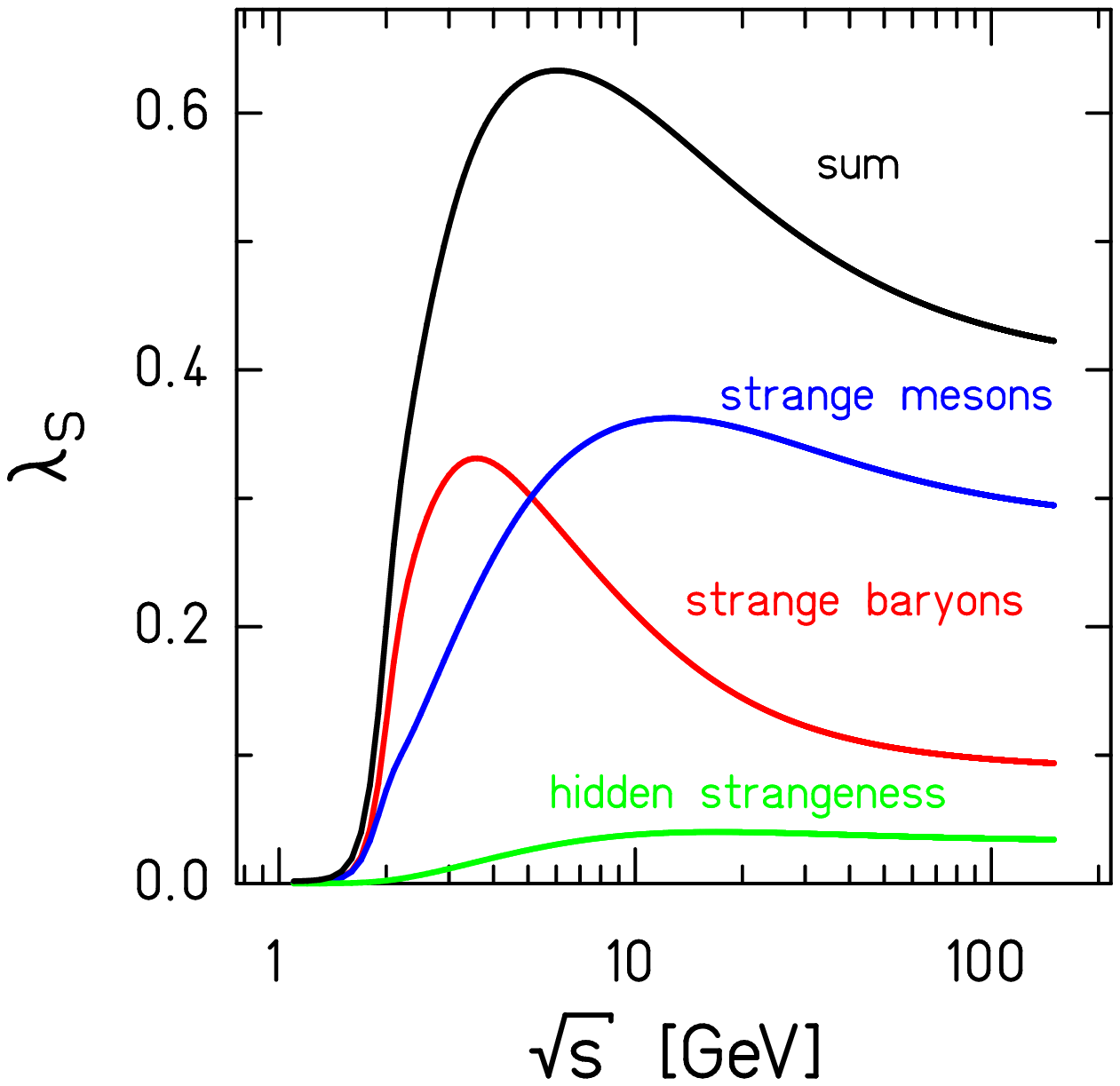}
\includegraphics*[width=7.5cm]{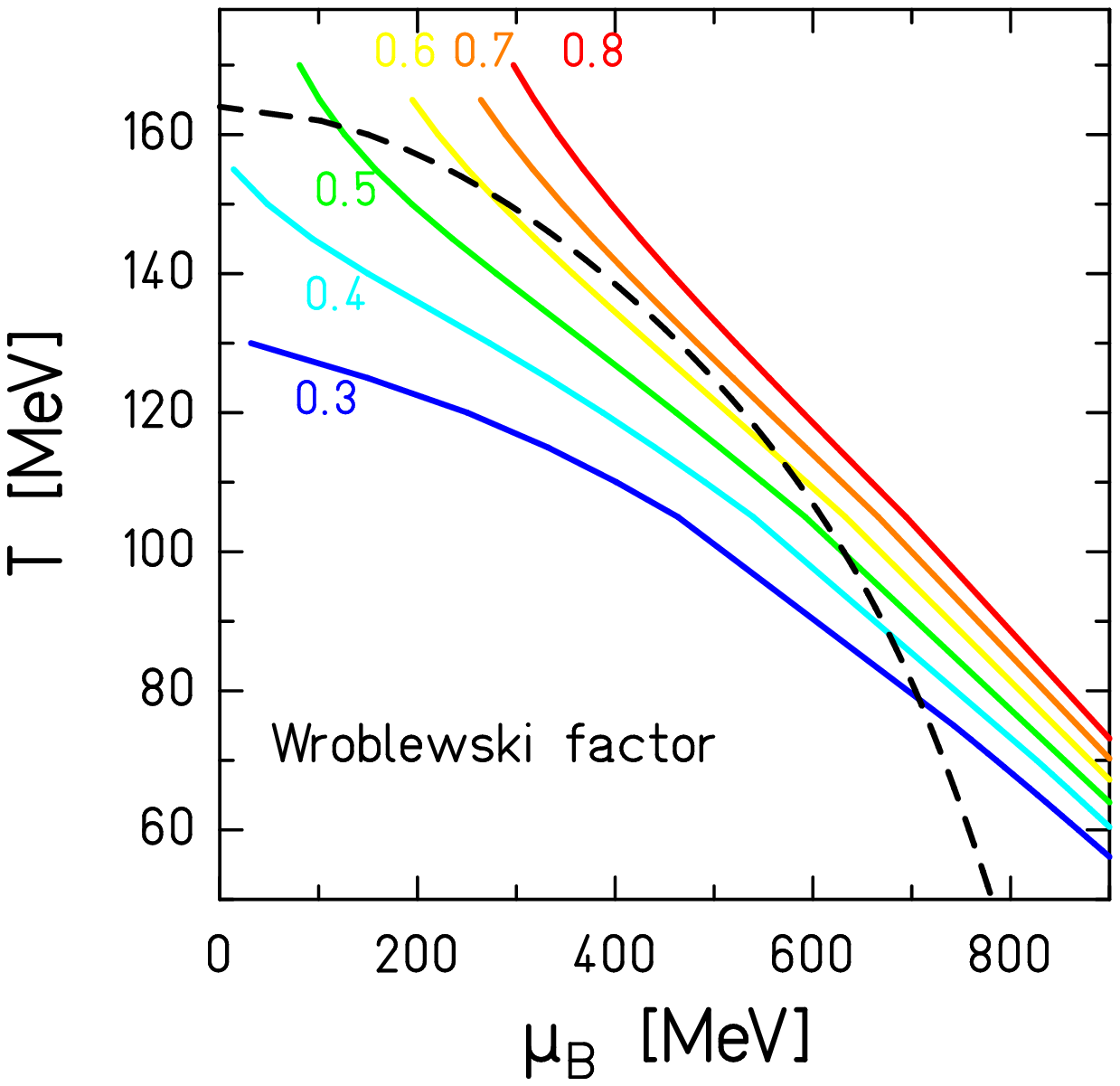}
 \caption{Left: Contributions to the Wroblewski
factor $\lambda_s$ (for definition see text) from strange baryons,
strange mesons, and mesons with hidden strangeness. The sum of all
contributions is given by the full black line. Right: Lines of
constant Wroblewski factor $\lambda_s$ in the $T-\mu_B$ plane
(solid lines) together with the freeze-out curve (dashed
line)~\protect\cite{1gev}.}
\end{center}
\label{Wrob}
\end{figure}

The solid line (marked ``sum'') in Fig.~1 describes the
Statistical-Model calculations in complete equilibrium along the
unified freeze-out curve~\cite{1gev} with the energy-dependent
parameters $T$ and $\mu_B$. From Fig.~1 we conclude that around
$\sqrt{s_{\rm NN}}$ = 8.2 GeV corresponding to an incident energy
of 30 $A$ GeV, the relative strangeness content in heavy-ion
collisions reaches a clear and well pronounced maximum. The
Wroblewski factor decreases towards higher energies and reaches a
limiting value of about 0.43. For details see
Ref.~\cite{max_strange}.

The appearance of the maximum can be traced  to the specific
dependence of $\mu_B$ and $T$ on the beam energy as also pointed
out in Ref.~\cite{SK}. Figure~1 (right) shows lines of constant
$\lambda_s$ in the $T-\mu_B$ plane. As expected, $\lambda_s$ rises
with increasing $T$ for fixed $\mu_B$. Following the chemical
freeze-out curve, shown as a dashed line in Fig.~1, one can see
that
 $\lambda_s$ rises quickly from SIS to AGS energies,
then reaches  a maximum at $\mu_B\approx 500$ MeV and $T\approx
130$ MeV. These freeze-out parameters correspond to 30 $A$ GeV
laboratory energy. At higher incident energies the increase in $T$
becomes negligible but $\mu_B$ keeps on decreasing and as a
consequence $\lambda_s$ also decreases.

The importance of finite baryon density on the behavior of
$\lambda_s$ is demonstrated in  Fig.~1 left, showing separately
the contributions to $\left<s\bar{s}\right>$ coming from strange
baryons, from strange mesons and from hidden strangeness,
i.e.~from hadrons  like $\phi$ and $\eta$. As can be seen in
Fig.~1, the origin of the maximum in the Wroblewski ratio can be
traced  to the contribution of strange baryons. The production of
strange baryons dominates at low $\sqrt{s_{\rm NN}}$ and loses
importance at high incident energies when the yield of strange
mesons increases. However, strange mesons also exhibit a broad
maximum. This is due to  associated production of e.g.~kaons
together with hyperons at the lower incident energies.\\

\begin{figure}[h]
\begin{center}
\vspace*{-1.0cm}
\includegraphics*[width=10cm]{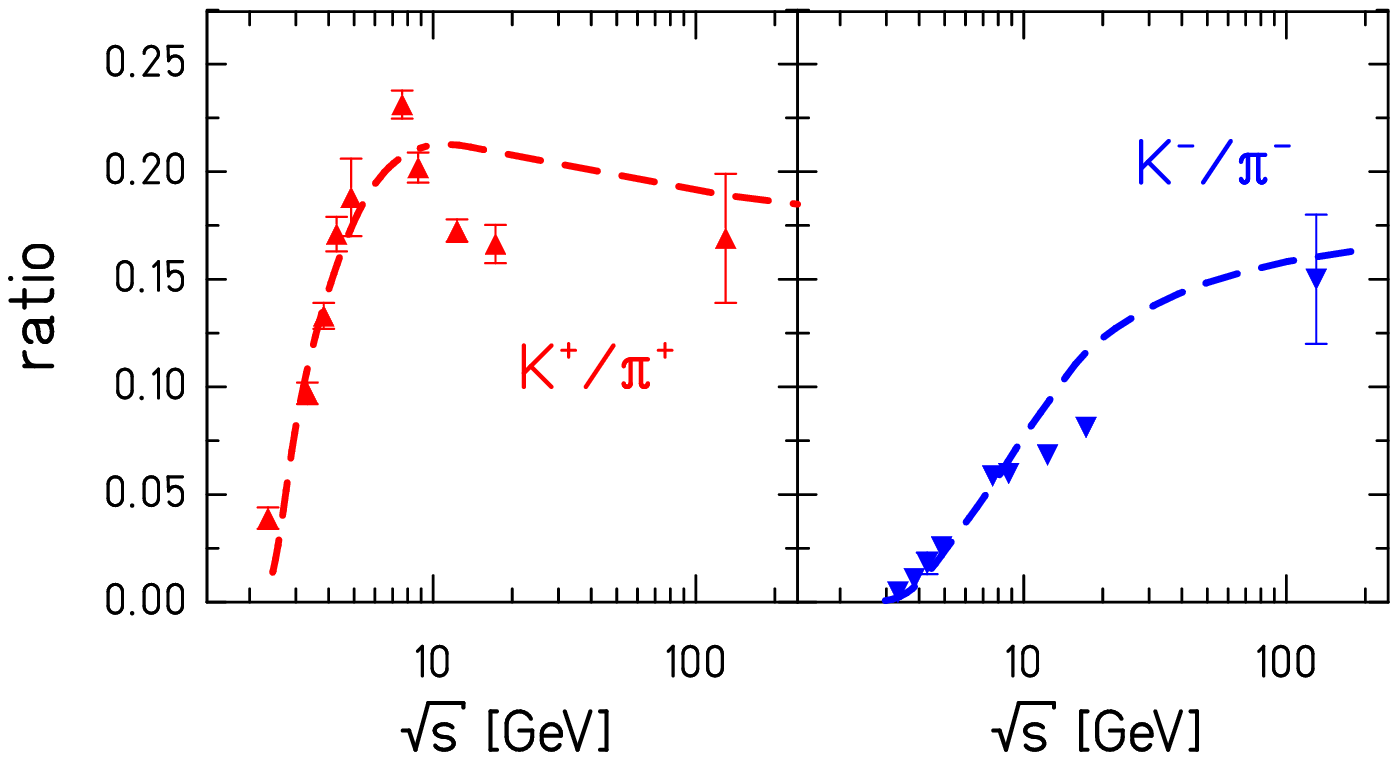}\\
\includegraphics*[width=13cm]{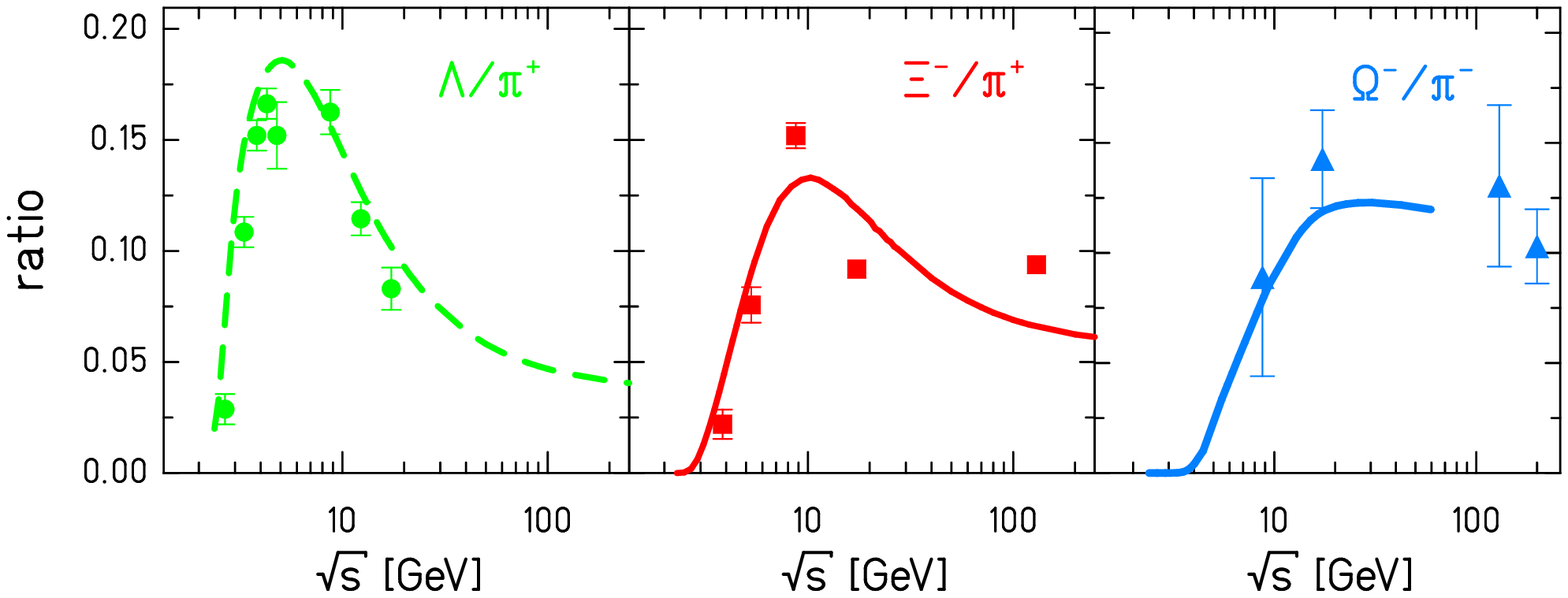}
\vspace*{-0.8cm}
\end{center}
\caption{Ratio of strange-to-non-strange mesons (upper part) and
 the corresponding ratios for baryons (lower part) as a function of $\sqrt{s_{\rm NN}}$.}
\label{Lambda_pi}
\end{figure}

Figure~\ref{Lambda_pi} shows  the comparison of the Statistical
Model~\cite{max_strange} and $4\pi$ data (except for RHIC data,
above $\sqrt{s_{\rm NN}}$ = 130 GeV). As can be understood from
the arguments above, the ratio $\Lambda/\pi$ exhibits the most
pronounced maximum, K$^+/\pi^+$ a weaker one and K$^-/\pi^-$ has
no maximum at all.

It is worth noting that the maxima in the
 ratios for multi-strange baryons occur at ever
higher beam energies. This can be seen clearly in
Fig.~\ref{Lambda_pi} for the $\Xi^-/\pi^+$ ratio which peaks at a
higher value of the beam energy. The ratio $\Omega^-/\pi^+$ also
shows a (very weak) maximum, as can be seen in
Fig.~\ref{Lambda_pi}.
 The higher the strangeness content of the baryon, the higher
in energy is the maximum. This behavior is due to a combination of
the facts that the baryon chemical potential decreases rapidly
with energy and the multi-strange baryons have successively higher
thresholds.

It is to be expected that if these maxima do not all occur at the
same temperature, i.e. at the same beam energy, then the case for
a phase transition is not very strong. The observed behavior seems
to be governed by properties of the hadron gas. More detailed
experimental studies of multi-strange hadrons will allow the
verification or disproval of the trends shown in this paper. It
should be clear that the $\Omega^-/\pi^+$ ratio is very broad and
shallow and it  will be difficult to find a maximum
experimentally.

In general, the model  gives a good description of the data. It
shows a broad maximum in the K$^+/\pi^+$ ratio while the data
exhibit a sharp peak. The drop towards 158 $A$ GeV is most
pronounced when using 4$\pi$ yields as done in these figures. This
decrease is less pronounced for midrapidity values~\cite{NA49}.

\section{Transition from baryonic to mesonic freeze-out}

While the Statistical Model cannot explain the sharpness of the
peak in the K$^+/\pi^+$ ratio, there are nevertheless several
phenomena giving rise to the rapid change which warrant a closer
look at the model.

\begin{figure}[h]
\begin{center}
\vspace*{-3.5cm}
\includegraphics[width=10.5cm]{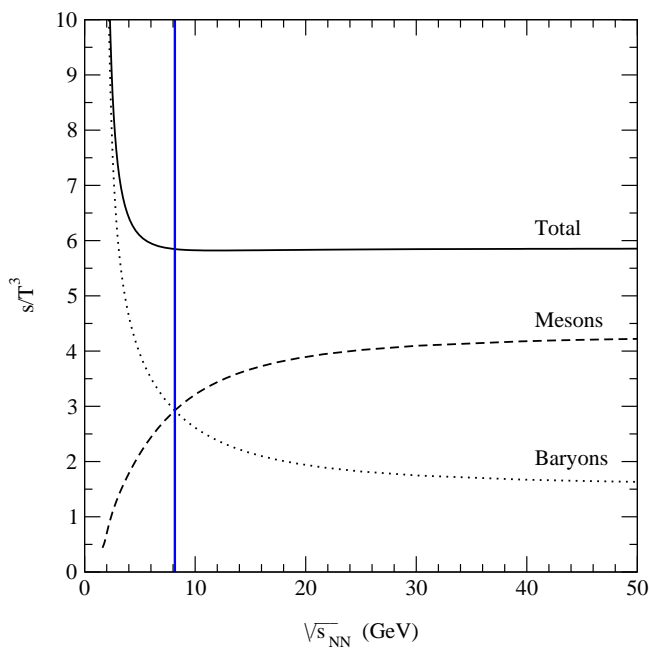}
\vspace*{-2.0cm}
 \caption{The entropy density normalized to $T^3$
as a function of the beam energy as calculated in the Statistical
Model using {\sc Thermus}~\cite{thermus}.} \end{center}
\label{entropy_s}
\end{figure}

To get a better estimate of the thermal parameters in the
transition region we show in Fig.~\ref{entropy_s} the entropy
density as a function of beam energy following the freeze-out
curve given in \cite{1gev}. The separate contribution of mesons
and of baryons to the total entropy is also shown in this figure.
There is a clear change of baryon to meson  dominance around
$\sqrt{s_{\rm NN}}$ = 8.2 GeV. Above this value the  entropy is
carried mainly by mesonic degrees of freedom. It is remarkable
that the entropy density divided by $T^3$ is constant for all
incident energies except for the  low-energy region corresponding
to the SIS energy region.

\section{Origin for a possible deviation from the freeze-out curve}

\begin{figure}[h]
\begin{center}
\includegraphics[width=9.3cm]{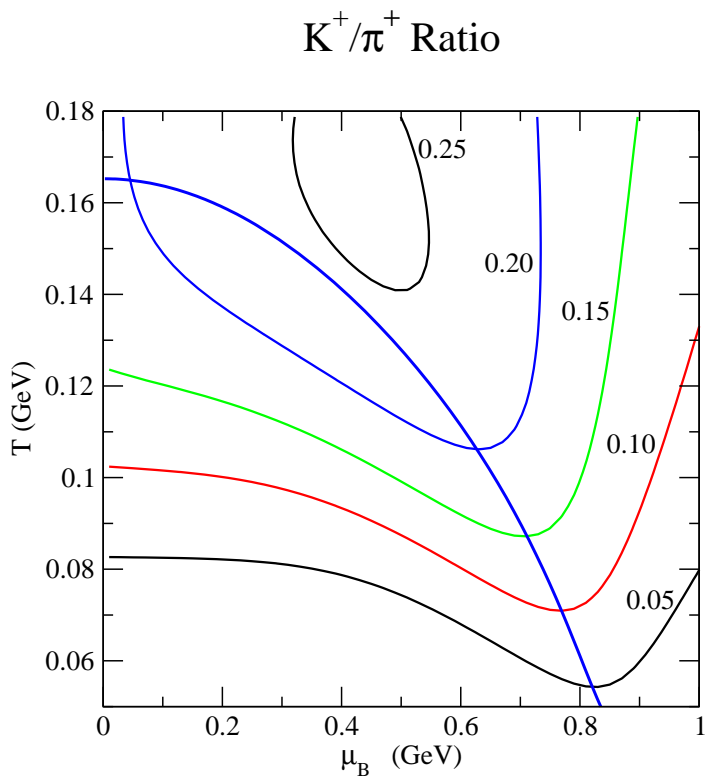}
 \caption{Values of the K$^+/\pi^+$ ratio for combinations
 of $T$ and $\mu_B$ are given by the contour lines and the corresponding values.
 The thick line refers to the freeze-out curve~\cite{1gev}.} \end{center}
\label{KP_PIP_T_mub}
\end{figure}

In this section, we explore the possibility that freeze-out might
happen earlier in the transition region. For this interpretation,
we show in Fig.~\ref{KP_PIP_T_mub} the calculated values of the
K$^+/\pi^+$ ratio for various combinations of $T$ and $\mu_B$ as
contour lines with the corresponding values given in the figure.
The thick solid line reflects the locations of the freeze-out
given by the condition of Ref.~\cite{1gev}. If freeze-out happens
around an incident energies of 30 $A$ GeV at higher $T$, then the
ratio K$^+/\pi^+$ will be higher. This ratio can never exceed a
value of 0.25 in an equilibrium condition.

It turns out that other particle ratios are less affected by a
different freeze-out scenario, as their variation in the $T-\mu_B$
plane is very different~\cite{SW}.

\begin{figure}[h]
\begin{center}
\includegraphics[width=9cm]{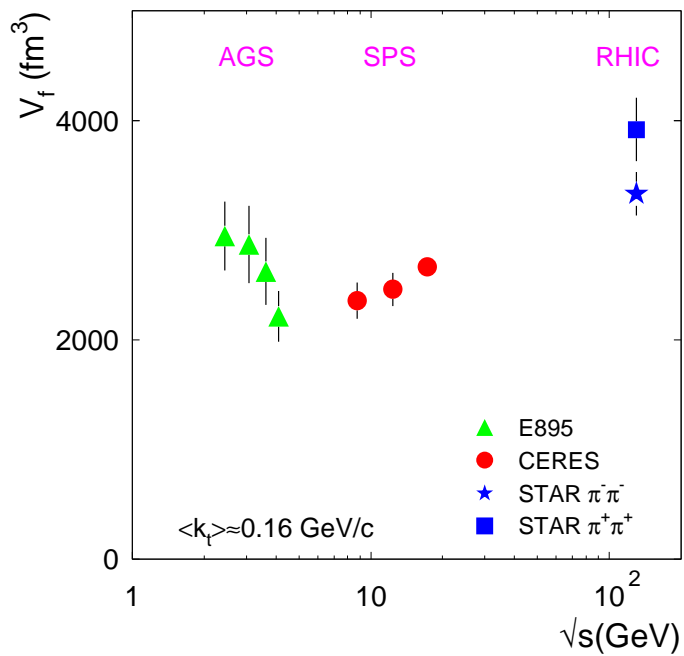}
 \caption{Freeze-out volumina as extracted from HBT studies~\cite{CERES}.}
 \end{center} \label{CERES}
\end{figure}

An early freeze-out is supported by results from HBT
studies~\cite{CERES}. Figure~5 shows the extracted volumina as a
function of $\sqrt{s_{\rm NN}}$. Between top AGS and the lowest
SPS energies a minimum can be seen. As the fireball is expanding,
a smaller volume reflects an earlier time. The authors of
Ref.~\cite{CERES} relate this minimum to a change in the
interaction from $\pi N$ to $\pi \pi$.

\section{Summary}

It has been shown that the Statistical Model yields a maximum in
the relative strangeness content around 30 $A$ GeV. This is due to
a saturation in the temperature $T$ while the chemical potential
keeps decreasing with incident energy. Since the chemical
potential plays a key role, it is clear that baryons are strongly
affected. Indeed, all hyperon/$\pi$ ratios yield maxima. In
contrast, the K$^-/\pi^-$ ratio shows a continuously rising curve
as expected from the arguments above. The K$^+/\pi^+$ ratio,
however, exhibits a maximum, as K$^+$ mesons are sensitive to the
baryo-chemical potential due to their associate production with
hyperons. The model predicts that for different hyperon/$\pi$
ratios the maxima occur at different energies. If experiments
prove this, the case for a phase transition is strongly weakened.

The energy regime around 30 $A$ GeV seems to have specific
properties. It is shown that the entropy production occurs below
this energy mainly via creation of baryons, while at the higher
incident energies meson production dominates.

In a third section, we speculated on the impact of a change in the
freeze-out condition which might lead to an early freeze-out, thus
deviating from the usual freeze-out condition. Such a scenario
would increase the K$^+/\pi^+$ while leaving other particle ratios
essentially unchanged. HBT studies show that around 30 $A$ GeV a
minimum in the extracted volumina occurs. This could be
interpreted as an earlier kinetic freeze-out and might indicate
also another freeze-out for chemical decoupling.\\

This work was supported by the German Ministerium f\"ur Bildung
und Forschung (BMBF) and by the Polish Committee of Scientific
Research.

\end{document}